\documentclass[11pt]{article}
\usepackage{moriond,epsfig,feynmf}
\newcommand{\Pf}{\ensuremath{\mathrm{f}}}
\newcommand{\Pl}{\ensuremath{\mathrm{l}}}

\newcommand{\Pe}{\ensuremath{\mathrm{e}}}

\newcommand{\Pq}{\ensuremath{\mathrm{q}}}

\newcommand{\Pcq}{\ensuremath{\mathrm{c}}}
\newcommand{\Pbq}{\ensuremath{\mathrm{b}}}

\newcommand{\Pg}{\ensuremath{\mathrm{g}}}
\newcommand{\PW}{\ensuremath{\mathrm{W}}}
\newcommand{\PZ}{\ensuremath{\mathrm{Z}}}

\newcommand{\grav}{\ensuremath{\mathrm{G}}}
\newcommand{\qq}{\ensuremath{\mathrm{q\bar{q}}}}

\newcommand{\ee}{\ensuremath{\mathrm{e^+e^-}}}

\newcommand{\pp}{\ensuremath{\mathrm{p\bar{p}}}}

\newcommand{\pbinv}{\ensuremath{\mathrm{pb^{-1}}}}
\newcommand{\fbinv}{\ensuremath{\mathrm{fb^{-1}}}}

\newcommand{\et}{\ensuremath{E_{\mathrm{t}}}}
\newcommand{\met}{\ensuremath{\not \!\! \et}}

\newcommand{\pt}{\ensuremath{p_{\perp}}}

\newcommand{\cha}{\ensuremath{\widetilde{\chi}}}

\newcommand{\snu}{\ensuremath{\tilde{\nu}}}

\newcommand{\stq}{\ensuremath{\mathrm{\tilde{t}}}}

\newcommand{\Ubar}{{\bar U}}
\newcommand{\Dbar}{{\bar D}}
\newcommand{\Ebar}{{\bar E}}
\newcommand{\lle}[3]{\ensuremath{L_{#1}L_{#2}\Ebar_{#3}}}
\newcommand{\lqd}[3]{\ensuremath{L_{#1}Q_{#2}\Dbar_{#3}}}
\newcommand{\udd}[3]{\ensuremath{\Ubar_{#1}\Dbar_{#2}\Dbar_{#3}}}

\newcommand{\iisoo}[9]{
   \fmfleftn{i}{2}\fmfrightn{o}{2}
   \fmflabel{#1}{i1}
   \fmflabel{#3}{i2}
   \fmflabel{#8}{o2}
   \fmflabel{#6}{o1}
   \fmf{#2}{i1,v1}
   \fmf{#4}{v1,i2}
   \fmf{#7}{v2,o1}
   \fmf{#9}{o2,v2}
   \fmf{boson,label=#5}{v1,v2}
   \fmfdot{v1,v2}
}
\newcommand{\iigoo}[9]{
   \fmfleftn{i}{2}\fmfrightn{o}{2}
   \fmflabel{#1}{i1}
   \fmflabel{#3}{i2}
   \fmflabel{#8}{o2}
   \fmflabel{#6}{o1}
   \fmf{#2}{i1,v1}
   \fmf{#4}{v1,i2}
   \fmf{#7}{v2,o1}
   \fmf{#9}{o2,v2}
   \fmf{zigzag,label=#5}{v1,v2}
   \fmfdot{v1,v2}
}

\def\Journal#1#2#3#4{{#1} {\bf #2}, #3 (#4)}

\def\PLB{{\em Phys. Lett.}  B}
\def\PRL{\em Phys. Rev. Lett.}
\def\PRD{{\em Phys. Rev.} D}

\def\EPJ{{\em Eur. Phys. J} C}

\def\be{\begin{equation}}
\def\ee{\end{equation}}
\def\bea{\begin{eqnarray}}
\def\eea{\end{eqnarray}}

\begin{document}

\begin{fmffile}{procfeyn}
\fmfset{arrow_len}{2.5mm}
\fmfset{arrow_ang}{13}
\fmfset{zigzag_len}{1mm}

\vspace*{4cm}
\title{SEARCHES FOR NEW PHYSICS AT THE TEVATRON}

\author{ V. B\"USCHER }

\address{Universit\"at Mainz, Institut f\"ur Physik, Staudinger Weg 7,\\
55099 Mainz, Germany}

\maketitle\abstracts{
Recent results of searches for new phenomena using data collected in
Run~I by the CDF and D0 experiments are presented. D0 set limits
for top squarks decaying into $\Pbq\Pl\snu$ significantly beyond the
reach of previous analyses in this channel ($m_{\stq}\!>\!140$~GeV at
95\%~C.L. for
$m_{\snu}\!=\!45$~GeV). CDF exclude top squarks decaying into $\Pbq\tau$ for 
masses below 111~GeV. In addition, D0 have analyzed photon and
electron pair production to search for effects of large extra
dimensions. No evidence is found, resulting in limits on the
effective Planck scale of 1.1 TeV.
Finally, D0 have used a model-independent search strategy for new
physics at high \pt\ to test more than 32 different final states for
evidence of a signal. Observations in all samples are found to be
consistent with expectations from the Standard Model. 
}

\section{Introduction}
With the end of Run~I in 1996 the two Tevatron experiments CDF and D0
each had collected about 110~\pbinv\ of \pp-collision data. Since then both
collaborations have been preparing upgrades of their detectors that
will allow to operate them during the next run 
of the upgraded Tevatron (Run~II). For Run~II, 3-4~\fbinv\ of data are
expected in the first few years of running.

The Run~I data have been analyzed extensively in search for new
physics by both experiments. New results from four analyses are
presented here. In addition, the sensitivity expected in these
channels for Run~II is discussed.

\section{Searches for Supersymmetry}
Supersymmetry predicts new supersymmetric degrees of freedom
for each Standard Model particle field. The scalar supersymmetric fields
associated with the left- and right-handed fermion fields can mix to
form mass eigenstates, potentially generating relatively light
scalar particles, in particular in the third generation. It is therefore
of interest to search for the production of light stop
quarks at the Tevatron. Both D0 and CDF have extended previous
analyses with new searches for top squarks: D0 are presenting a new
search within the Minimal Supersymmetric Standard Model (MSSM), CDF
have analyzed their data in search for top squarks in Supersymmetry
with R-parity violation.

In \pp-collisions top squarks are pair-produced strongly via
\qq-annihilation or gluon-gluon-fusion. Cross-sections have been
calculated to next-to-leading order~\cite{stop_nlo} and are of the order of
10~pb for $m_{\stq}=100$~GeV.

\subsection{Top Squarks in the MSSM}
Within the MSSM, top squarks that are light enough to be observed at
the Tevatron decay primarily into $\Pbq\cha^+$ (if
$m_{\stq}>m_{\Pbq}+m_{\cha^+}$), $\Pbq\Pf\Pf\chi^0_1$, $\Pcq\chi^0_1$
or $\Pbq\Pl\snu$. 
The latter decay mode is dominant for light sneutrinos and would lead
to final states containing two b-jets, two leptons and missing
transverse energy from sneutrinos escaping detection.
D0 have searched for pair production of top squarks in this channel in
final states containing one electron and one muon, as backgrounds in
this mode are particularly small. 

Based on a previous analysis~\cite{d0_emux} of the final state
$\Pe\mu\met+X$, events are selected if they contain one isolated
electron and muon with $\pt\!>\!15$~GeV and missing transverse energy of at
least 20~GeV. The electron and muon are required to be central and acoplanar to
suppress background from $\PZ\to\tau^+\tau^-$ and QCD. After all cuts,
$13.4\pm 1.5$ events are expected within the Standard Model. Typical
signal efficiencies range from 0.5\% ($m_{\stq}\!=\!100$~GeV,
$m_{\snu}\!=\!70$~GeV) to 4.0\% ($m_{\stq}\!=\!140$~GeV,
$m_{\snu}\!=\!40$~GeV).

\begin{figure}
\epsfig{figure=xsec.tautau.epsi,height=6.6cm}\hfill
\epsfig{figure=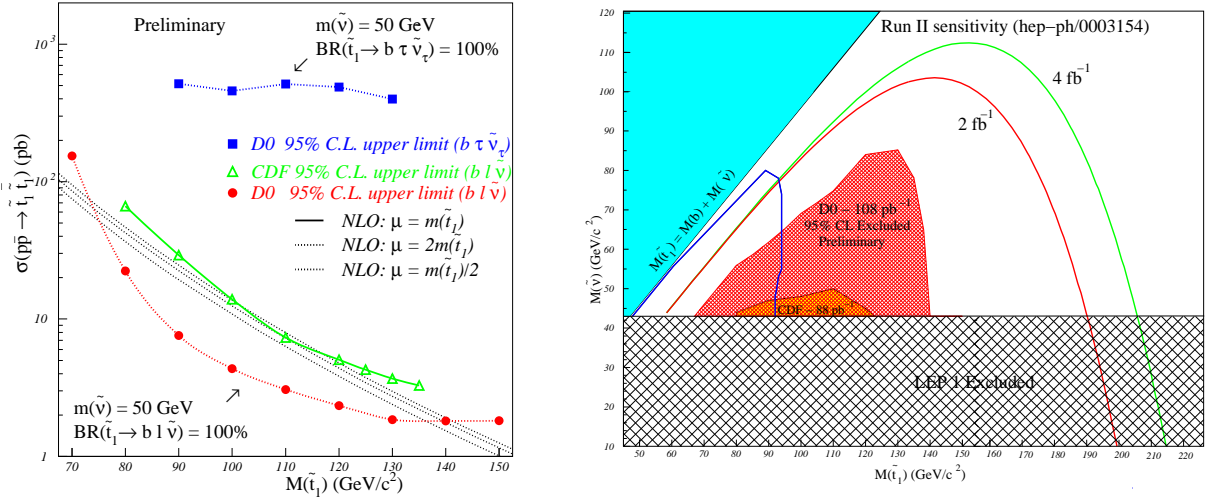,height=6.6cm}\\
\caption{Limits on the pair production cross section of top squarks
decaying into $\Pbq\Pl\snu$ with $m_{\snu}\!=\!50$~GeV in comparison
with the MSSM expectation (left); Regions in $m_{\snu}$ versus $m_{\stq}$
excluded by D0 (this analysis), CDF~\protect\cite{stop_snu_cdf} and
LEP~\protect\cite{stop_snu_lep}
(right). The sensitivity expected for Tevatron
Run~II~\protect\cite{SUGRA_report} is 
plotted for integrated luminosities of 2~\fbinv\ and 4~\fbinv.
\label{fig:stop_limits}}
\end{figure}

In the D0 data 11 events are selected and therefore no evidence of signal
is observed. From this, upper limits on the top squark pair
production cross section have been derived at
95\%~C.L. (Fig.~\ref{fig:stop_limits}).
Assuming
$\mathrm{BR}(\stq\to\Pl\Pbq\snu)\!=\!1$, this limit corresponds to an
excluded region in 
the $(m_{\stq},m_{\snu})$-plane as indicated in
Fig.~\ref{fig:stop_limits}, significantly improving on existing limits 
from LEP~\cite{stop_snu_lep} and CDF~\cite{stop_snu_cdf}.

The expected sensitivity for Run~II has been estimated for this
channel using a MC study~\cite{SUGRA_report}. Due to the high
luminosity at an 
increased center of mass energy (2~TeV), the reach of this analysis
will be extended considerably beyond the limit presented here: top squark
masses of 200~GeV and beyond will be accessible for sneutrino masses
around 50~GeV (Fig.~\ref{fig:stop_limits}).

\subsection{Top Squarks with R-Parity Violation}
The most general superpotential~W contains a number of terms that lead
to violation of lepton- or baryon-number and that are generally removed by
imposing the conservation of R-parity:
\begin{eqnarray}
\nonumber
W & = &   W_{RPC} + W_{TRPV} + W_{BRPV}\\
\nonumber
  & = &   \bar{U} y_u Q H_u - \bar{D} y_d Q H_d - \bar{E} y_e L H_d +
 \mu H_u H_d +\\
\nonumber
   & & \lambda_{ijk}\lle{i}{j}{k}+
               \lambda'_{ijk}\lqd{i}{j}{k}+
               \lambda''_{ijk}\udd{i}{j}{k}+\\
   & & \epsilon_i L_i H_u
\label{eq:rpv}
\end{eqnarray}
Here $U,D,Q$ and $L,E$ denote the quark and lepton superfields,
$\mu$ is the Higgsino mass parameter and $y$ contain the Yukawa
couplings. The first 4 terms ($W_{RPC}$) do not violate R-parity 
conservation and generate the standard MSSM lagrangian. In addition
there are 45 trilinear couplings $\lambda_{ijk}$ and 3 bilinear
couplings $\epsilon_i$ (with generation indices i,j,k).

Both trilinear ($\lambda'_{333}$) and bilinear ($\epsilon_3$)
terms can generate a tree-level 
coupling between top squark, b quark and $\tau$ lepton. Therefore,
when considering models without the conservation of R-parity, the
additional decay mode $\stq\to\Pbq\tau$ has to be taken into account
and can even dominate if the decay into chargino is kinematically not
accessible. 

\begin{figure}
\epsfig{figure=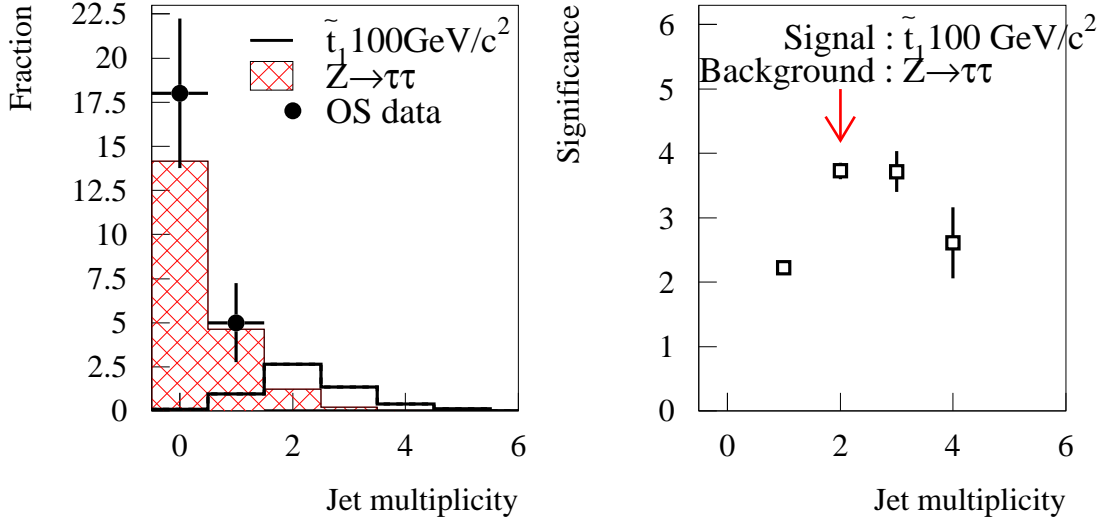,width=\textwidth}
\caption{Number of reconstructed jets in the search for
$\stq\to\Pbq\tau$ for the signal, background from
$\PZ\to\tau\tau$ and data (left); The expected significance of a top
squark signal is maximized by requiring at least two jets (right).
\label{fig:stop_rpv_njets}}
\end{figure}
Top squark pair production in this scenario leads to final states
containing two tau leptons and two b quarks. This signature has been 
considered in a previous search for third generation
leptoquarks~\cite{LQ3_Tevatron}, requiring one tau to decay
hadronically, the other 
leptonically to an electron. The CDF collaboration has reoptimized
their previous analysis 
after improvements in the identification of hadronic tau decays, which
has been enhanced by using tracking information as well as
$\pi_0$-reconstruction.
To minimize the dependency on the MC modelling of the tau
identification (as well as to cancel other systematic errors), the
number of signal-like events is measured relative 
to a measurement of the $\PZ\to\tau\tau$ cross section. Rejecting
events with one or more jets in the final state, $48\pm8$ events from
$\PZ\to\tau\tau$ are selected in the data. In contrast to this, two
jets or more are required to select events from top squark pair
production (Fig.~\ref{fig:stop_rpv_njets}).
This and other signal cuts have been chosen to maximize the
significance for $m_{\stq}=100$~GeV. 

\begin{figure}
\begin{center}
\epsfig{figure=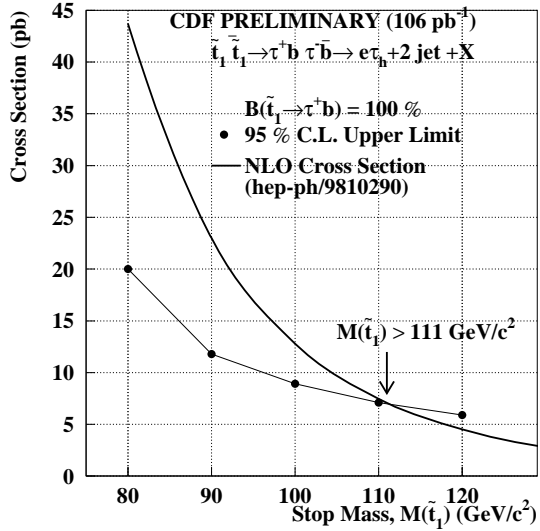,width=0.5\textwidth}
\caption{Limit on the top squark pair production cross section for
$\stq\to\Pbq\tau$ in comparison with the cross section expected within 
the MSSM. Top squark masses below 111~GeV are excluded at 95\%~C.L.
\label{fig:stop_rpv_limit}}
\end{center}
\end{figure}

After all cuts signal events are selected with an efficiency of 1.6\%
for $m_{\stq}=100$~GeV.
No events are observed in the data, consistent with the expectation of
$1.91\pm0.11(stat)\pm0.15(syst)$ from Standard Model backgrounds. 
Fig.~\ref{fig:stop_rpv_limit} shows the limit on the top squark pair
production cross section derived from this observation after
normalization to the $\PZ\to\tau\tau$ cross section. Based on the
NLO cross section calculation, top squarks lighter than 111~GeV can be
excluded at 95\%~C.L. for $\mathrm{BR}(\stq\to\Pbq\tau)\!=\!1$.

\section{Search for large extra dimensions}
\begin{figure}[b]
\begin{center}
\vspace{0.4cm}
\begin{fmfgraph*}(80,60)
\iisoo{$\Pq$}{fermion}{$\bar{\Pq}$}{fermion}{$\gamma,,\PZ$}
         {$\Pl^-$}{fermion}{$\Pl^+$}{fermion}
\end{fmfgraph*}
\hspace{1cm}
\begin{fmfgraph*}(80,60)
\iigoo{$\Pq$}{fermion}{$\bar\Pq$}{fermion}{$\grav_n$}
         {$\Pl^-$}{fermion}{$\Pl^+$}{fermion}
\end{fmfgraph*}
\hspace{1cm}
\begin{fmfgraph*}(80,60)
\iigoo{$\Pg$}{gluon}{$\Pg$}{gluon}{$\grav_n$}
         {$\Pl^-$}{fermion}{$\Pl^+$}{fermion}
\end{fmfgraph*}
\end{center}
\caption{Feynman diagrams contributing to dilepton production in
\pp-collisions for models with gravitons~$G_n$ allowed to propagate in
large extra dimensions. 
\label{fig:led_graphs}}
\end{figure}
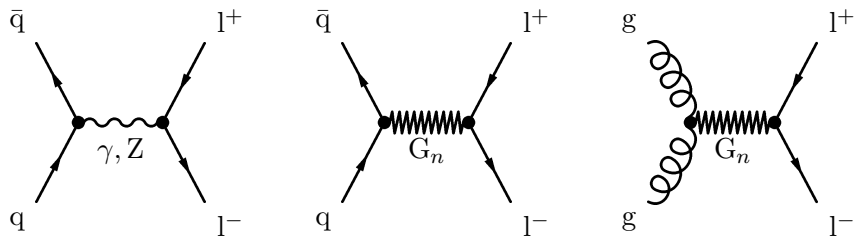

In models with large extra dimensions~\cite{LED_ADD}, the effective Planck
scale~$M_S$ can be as low as $O(1\,\mathrm{TeV})$, which would be close
enough to the weak scale to have considerable impact on the hierarchy problem. 
Effects from large extra dimensions can be observed at high energy
colliders via direct production of gravitons as well as via anomalous
difermion or diboson production proceeding through virtual graviton exchange
(Fig.~\ref{fig:led_graphs}). 

\begin{figure}[t]
\begin{center}
\epsfig{figure=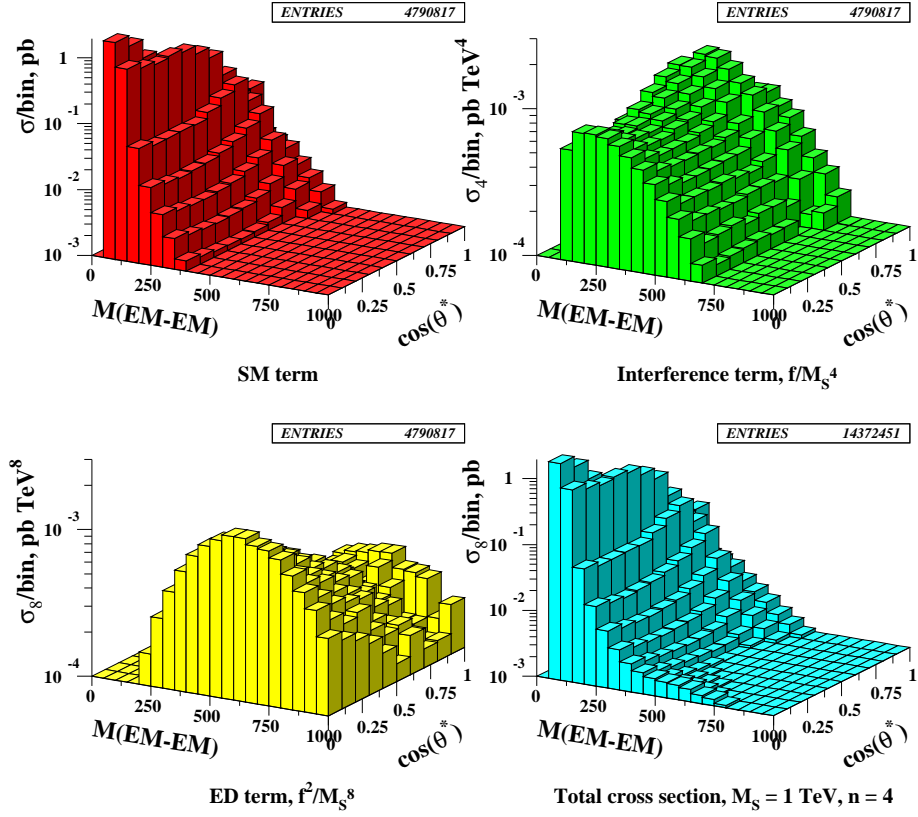,width=0.8\textwidth}
\caption{Di-EM cross section as a function of di-EM mass~M and
production angle $\theta^*$ for the Standard Model term (top left),
interence term (top right), Kaluza-Klein term (bottom left) as well as 
the sum of all terms (bottom right).
\label{fig:led_method}}
\end{center}
\end{figure}

D0 have searched for virtual graviton effects in dielectron and
diphoton production, taking into account both the invariant mass~M and the
production angle~$\cos\theta$ of the system. The differential cross-section
in the presence of large extra dimensions is modified by contributions
from Kaluza-Klein towers of graviton states and can be parametrized as
\begin{equation}
\frac{d^2\sigma}{d \cos\theta^* d M}=\frac{d^2\sigma_{SM}}{d
\cos\theta^* d M} + \eta_G f_{int}(\cos\theta^*,M)+
\eta_G^2 f_{KK}(\cos\theta^*,M),
\label{eq:led}
\end{equation}
with an interference term $f_{int}$, a pure Kaluza-Klein term
$f_{KK}$ and a model-dependent constant $\eta_G$ containing the
dependence on $M_S$ and the number of extra
dimensions. Fig.~\ref{fig:led_method} 
demonstrates the difference in shape comparing both $f_{int}$ and
$f_{KK}$ to the Standard Model cross section.

Since the Standard Model background from misidentified electrons or
photons at high invariant masses is negligible, very loose
identification cuts have been applied in this analysis. In particular, no
tracking information is used, thereby effectively treating both
photons and electrons as EM objects. Requiring two isolated EM
objects, about 80\% of the signal events pass the
selection. Efficiencies are calculated with Graviton-induced effects as
implemented in a \mbox{MC} generator at leading order. Higher order effects
are partially modelled by adding a transverse momentum to the di-EM system,
based on the transverse momentum spectrum of di-EM events observed in
the data.

\begin{figure}
\epsfig{figure=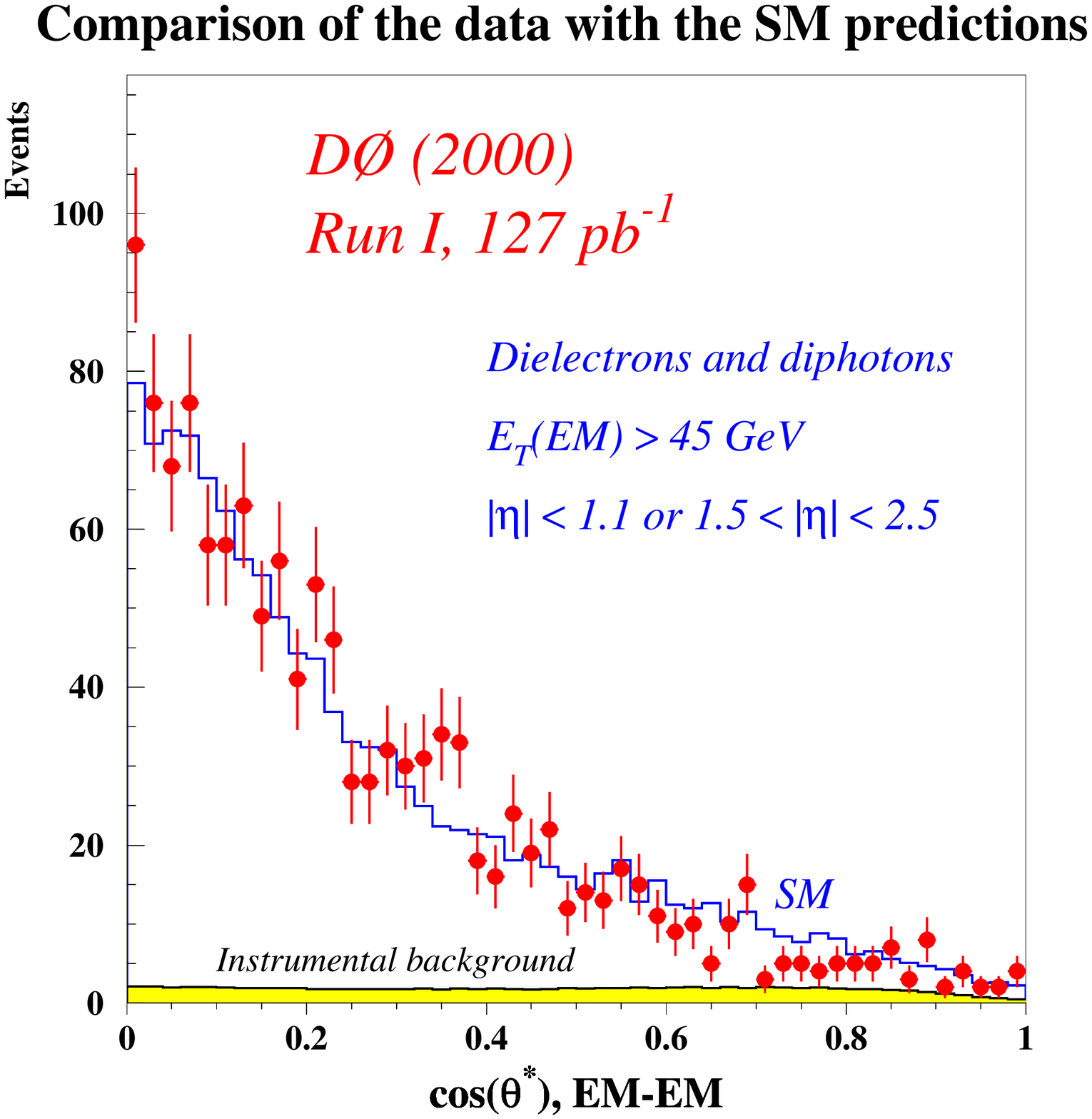,width=0.49\textwidth}\hfill
\epsfig{figure=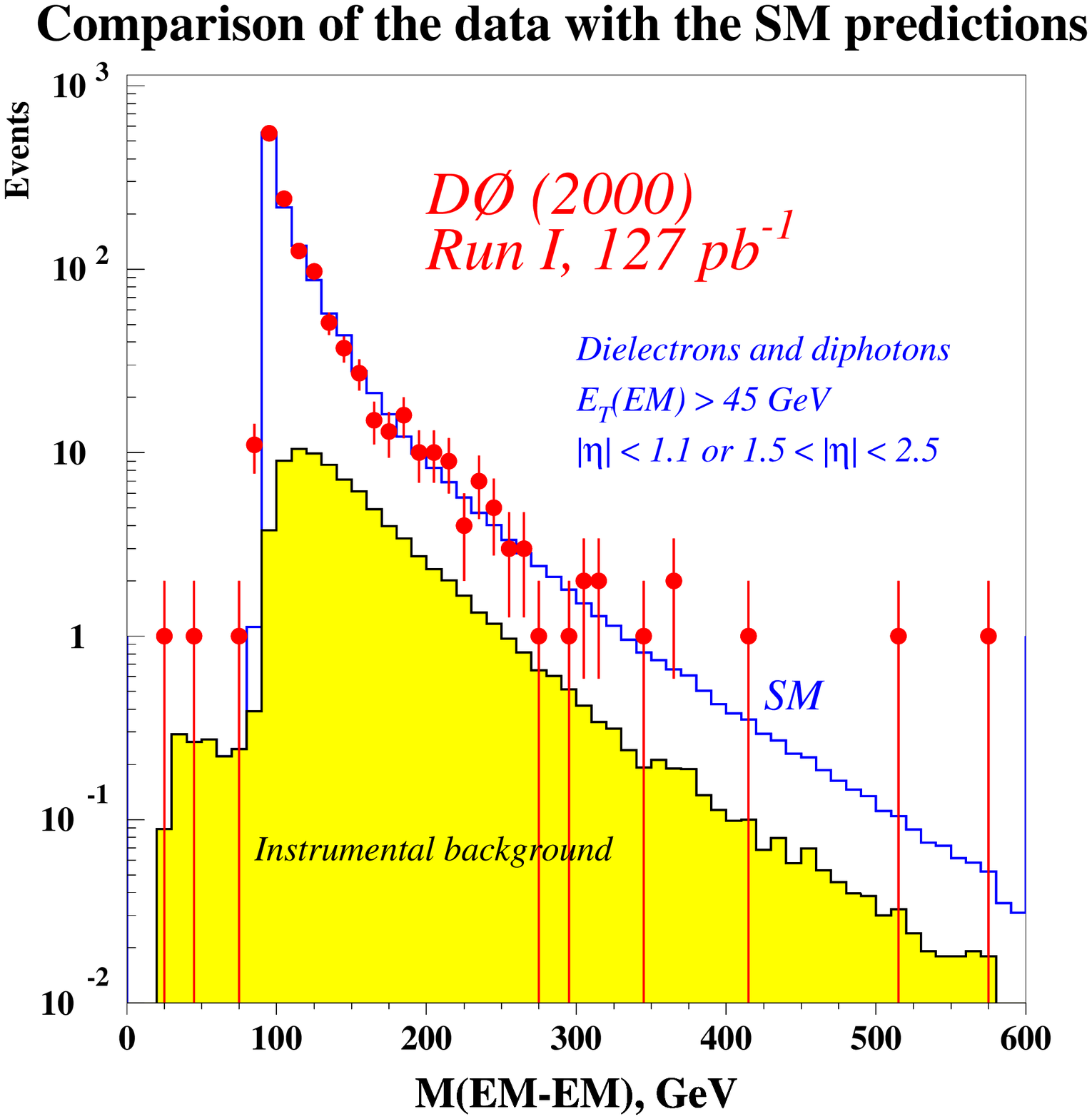,width=0.49\textwidth}\\
\caption{Distribution of di-EM production angle~$\theta^*$ (left) and
di-EM invariant mass~M (right) for D0 data (dots) and expectations
from the Standard Model (open histogram); Contributions from
instrumental backgrounds are displayed as the solid histogram.
\label{fig:led_data}}
\end{figure}

Backgrounds are entirely dominated by 
Drell-Yan and direct diphoton production, with only small
contributions from instrumental backgrounds
(Fig.~\ref{fig:led_data}).
The data are consistent with the expectations for di-EM production from
Standard Model sources throughout the entire
($M,\cos\theta$)-plane. Therefore, a fit to the data of
Equation~\ref{eq:led} yields a value for $\eta_G$ consistent with
zero, limiting the contribution from virtual graviton corrections to
$\eta_G\!<\!0.46\,\mathrm{TeV}^{-4}$ at 95\%~C.L. This result can be
translated into a 
lower limit on the effective Planck scale within a particular model of 
large extra dimensions. For instance, using the formalism introduced
by Hewett~\cite{LED_Hewett}, $M_S$ is found to be larger than 1.1~TeV at
95\%~C.L., comparable to current limits from LEP2~\cite{LED_LEP2}.
Extrapolating the D0 results to the Tevatron Run~II~\cite{LED_Run2},
effective Planck masses in the range of 1.5 to 2.5~TeV are expected to 
be probed with integrated luminosities of the order of 2~\fbinv.

\section{Model-independent Searches}
Both CDF and D0 have published searches for a large number of final
states and numerous models of new physics beyond the Standard
Model. While this is aimed at an optimal sensitivity for the
particular models 
considered in these searches, a more systematic approach seems
appropriate to ensure that no signal has been missed. In this context
D0 have developed a method to 
automatically find a significant excess of data at high \pt\ with
respect to the Standard Model expectation~\cite{Sleuth_method}. This
method has now been applied to 32 different final states to test the
full D0 Run~I data sample for hints of physics beyond the Standard Model.

Briefly, the method proceeds in the following steps (for details the
reader is referred to~\cite{Sleuth_method}): after splitting the dataset into
various exclusive final states, the d-dimensional space of variables
to be considered is mapped into a d-dimensional unit-box such that the
background distribution is flat. Regions of interest at high \pt\ are
defined in the unit-box, and for each region, the probability~R that
the expected background fluctuates to or above the number of events seen in
the data is calculated. After scanning all possible regions for small
values of R, the significance P of observing values as small as the
minimum R is calculated using a set of MC experiments that are
subjected to the exact same procedure as the data.

The final states considered in the D0 data include: $\Pe\mu\mathrm{X}$,
$\PW+\mathrm{jets}$, $\Pe\!\met+\mathrm{jets}$,
$\PZ+\mathrm{jets}$, $\Pe\Pe+\mathrm{jets}$, $\mu\mu+\mathrm{jets}$,
$\Pl/\gamma$ $\Pl/\gamma$ $\Pl/\gamma$ X, $\PW\gamma$ as well as
dijets. Standard and well-understood criteria are used for
reconstruction and identification of electrons, photons, muons, jets and
\met. Depending on the final state, the variables used in the
method are \met, the \pt\ of reconstructed leptons, jets and vector
bosons (W and Z bosons are reconstructed from leptons and the missing
\et\ vector).

\begin{figure}
\epsfig{figure=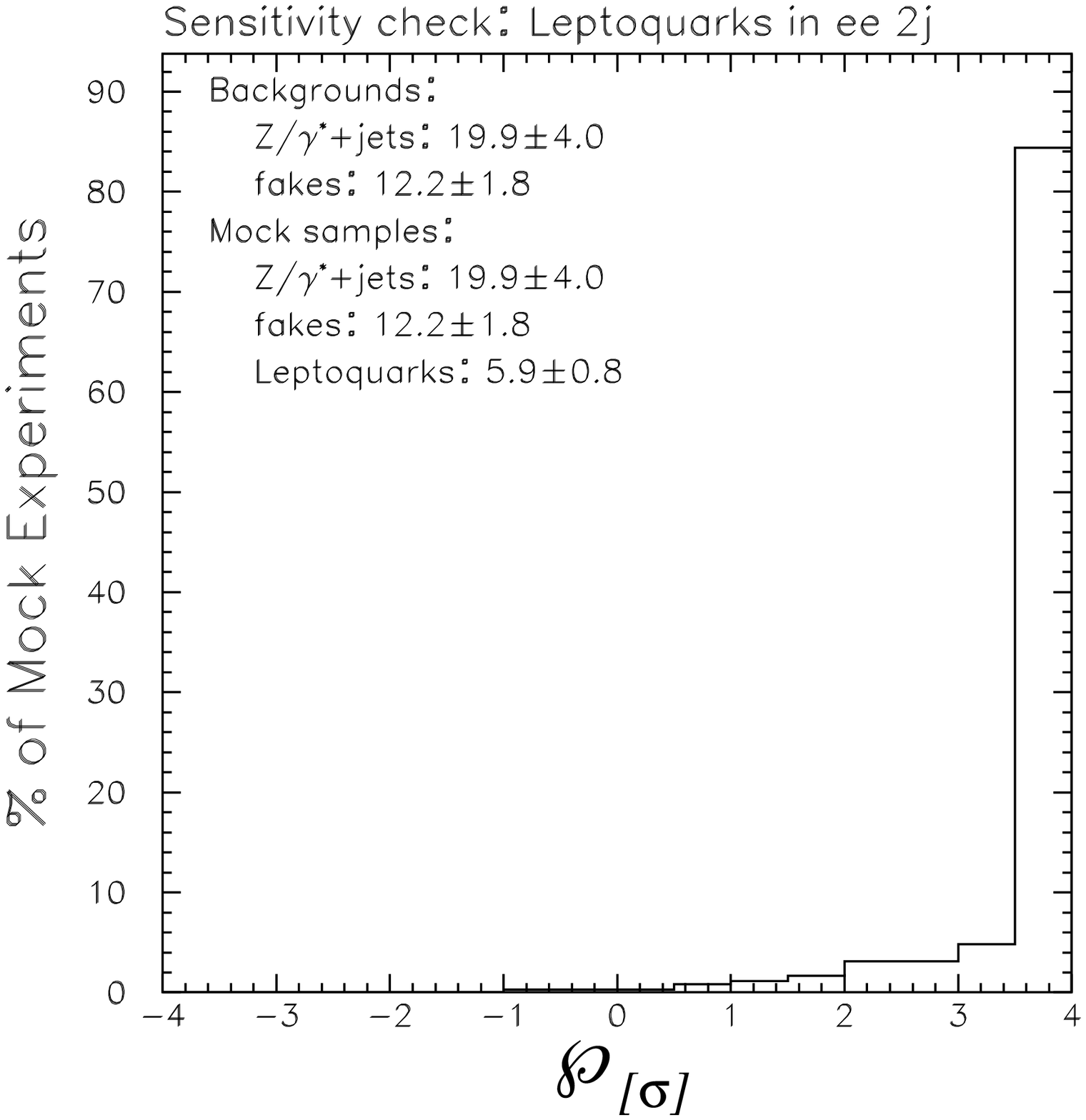,width=0.49\textwidth}\hfill
\epsfig{figure=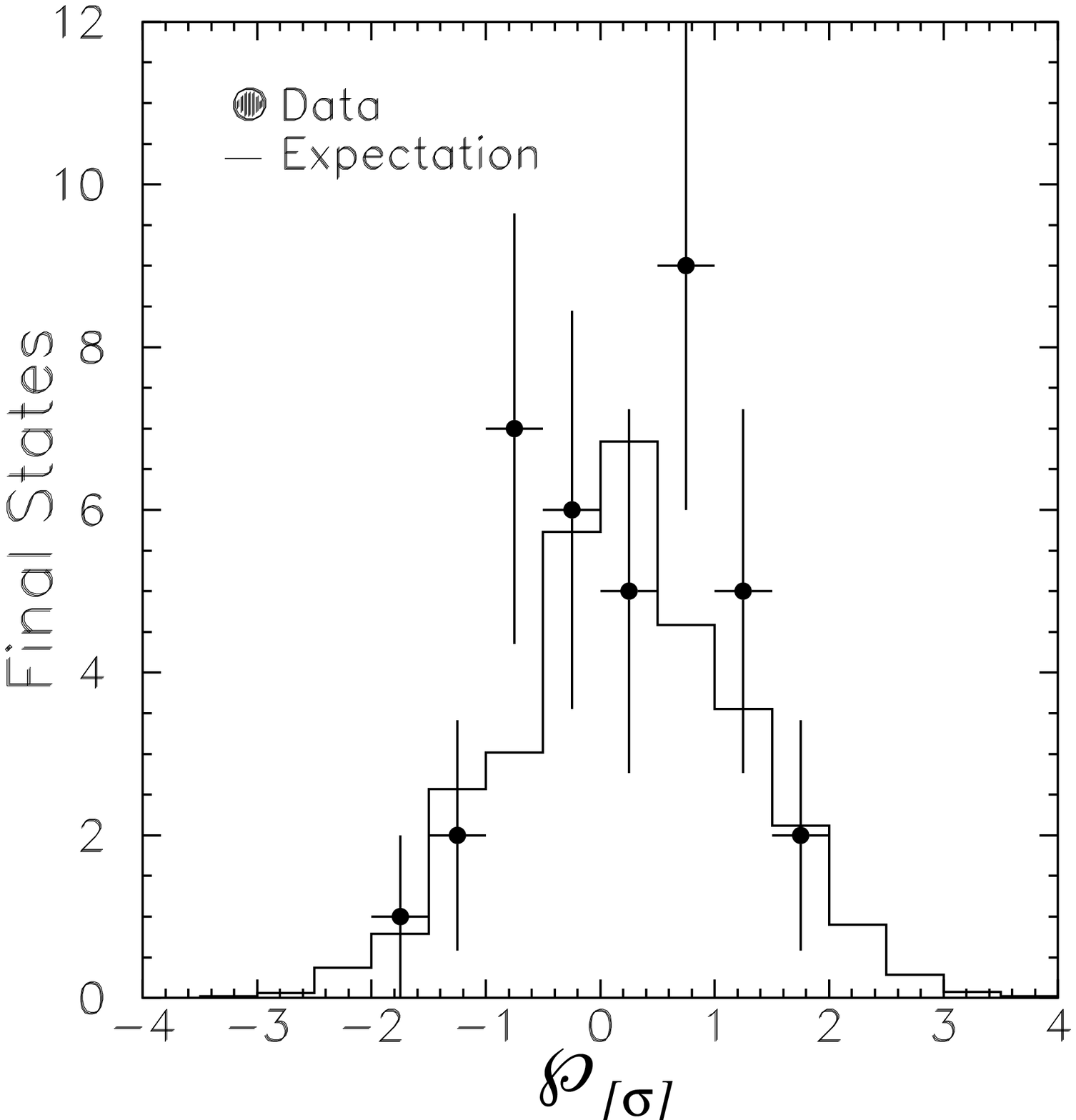,width=0.49\textwidth}\\
\caption{Distribution of significance~P (see text) for MC experiments
containing a signal from leptoquark pair production (left) and for all 
final states considered in D0 data in comparison with expectations
from Standard Model MC (right).
\label{fig:sleuth_plots}}
\end{figure}

To check whether the method is sensitive to a high-\pt\ signal in the
data, fake data samples have been generated adding events from pair
production of first generation leptoquarks
($m_{LQ}\!=\!170$~GeV). Fig.~\ref{fig:sleuth_plots} shows the distribution
of the significance P as measured in 100 such MC experiments. In about
90\% of the experiments evidence for signal is found at a significance 
of more than $3\sigma$.

Applying the method to the D0 Run~I dataset, the significance~P is
calculated for the full list of final states described
above. Fig.~\ref{fig:sleuth_plots} shows the distribution of P in
comparison with the expectation from background MC experiments. The
data are in good agreement with the Standard Model, therefore no
evidence for a significant excess has been observed.

\section{Summary}
Recent results of searches for new phenomena using data collected in
Run~I by the CDF and D0 experiments have been presented. Both
collaborations have expanded 
their searches for evidence of supersymmetry with new analyses on the
production of top squarks. D0 set a limit for top squarks decaying into 
$\Pbq\Pl\snu$ significantly beyond the reach of previous analyses in
this channel (e.g. $m_{\stq}\!>\!140$~GeV at 95\%~C.L. for
$m_{\snu}\!=\!45$~GeV). CDF 
exclude top squarks decaying into $\Pbq\tau$ for 
masses below 111~GeV.
In addition, D0 have analyzed photon and
electron pair production to search for effects of large extra
dimensions. No evidence was found, resulting in limits on the
effective Planck scale of 1.1~TeV.
Finally, D0 have used a model-independent search strategy for new
physics at high \pt\ to test more than 32 different final states for
evidence of a signal. Observations in all samples are found to be
consistent with expectations from the Standard Model.

\section*{Acknowledgments}
The author would like to thank the CDF and D0 contacts for working
hard to make information and preliminary results available in time for
the Moriond conference.

\end{fmffile}
\end{document}